\documentclass{iopconfser}
\usepackage{newtxmath}

\usepackage{graphicx}
\usepackage{amsmath}
\usepackage{tensor}
\usepackage{amssymb}
\usepackage{multicol}
\usepackage{sfmath}
\usepackage{pifont}
\usepackage[dvipsnames]{xcolor}
\usepackage{orcidlink}
\usepackage{acronym}
\definecolor{BoxCol}{rgb}{0.9,0.9,0.9}

\definecolor{BoxCol}{rgb}{0.9,0.9,1}

\definecolor{SectionCol}{rgb}{0,0,0.5}

\newcommand{\un}[1]{\mathsf{\,#1}}

\newcommand{\Porb}{P_{\text{orb}}}
\newcommand{\Tasc}{t_{\text{asc}}}
\newcommand{\TascThen}{\Tasc}
\newcommand{\TascNow}{\Tasc'}

\newcommand{\Tmax}{T_{\text{max}}}

\newcommand{\dcc}{LIGO-P2500519-v4}

\acrodef{NS}[NS]{neutron star}
\acrodef{BH}[BH]{black hole}
\acrodef{GW}[GW]{gravitational wave}
\acrodef{LMXB}[LMXB]{low-mass X-ray binary}
\acrodef{BBH}[BBH]{binary black hole}
\acrodefplural{LMXB}[LMXBs]{low-mass X-ray binaries}
\acrodef{AMXP}[AMXP]{accreting millisecond X-ray pulsar}
\acrodef{CW}[CW]{continuous gravitational wave}
\acrodef{HW}[HW]{hardware}
\acrodef{SW}[SW]{software}
\acrodef{PSD}[PSD]{power spectral density}
\acrodef{SNR}[SNR]{signal-to-noise ratio}
\acrodef{CPU}[CPU]{central processing unit}
\acrodef{SFT}[SFT]{short Fourier transform}
\acrodef{LLO}[LLO]{LIGO Livingston Observatory}
\acrodef{LHO}[LHO]{LIGO Hanford Observatory}
\acrodef{IFO}[IFO]{interferometer}
\acrodef{GPS}[GPS]{Global Positioning System}
\acrodef{ScoX1}[Sco~X-1]{Scorpius~X-1}
\acrodef{EoS}[EoS]{equation of state}
\acrodef{SSB}[SSB]{Solar system barycenter}
\acrodefplural{EoS}{equations of state}

\begin{document}

\title{Investigating Hardware Injections in LIGO O3
	Data: Simulated Signals from a Neutron Star in a
	Low-Mass X-ray Binary}

      \author{Jediah Tau\,\orcidlink{0009-0004-7428-762X}$^{1,2}$
        and John T. Whelan\,\orcidlink{0000-0001-5710-6576}$^{1,3}$}

\affil{$^1$Center for Computational Relativity and Gravitation,
  Rochester Institute of Technology, Rochester, NY 14623, USA}
\affil{$^2$School of Physics and Astronomy, Rochester Institute of
  Technology, Rochester, NY 14623, USA}
\affil{$^3$School of Mathematics and Statistics, Rochester Institute
  of Technology, Rochester, NY 14623, USA}

\email{jgt7263@rit.edu}

\begin{abstract}
  During LIGO-Virgo-KAGRA's third observing run (O3), simulated \ac{CW} signals
  were added to the LIGO detectors as \ac{HW} injections, which are
  physical injections into the interferometer control loop via auxiliary laser actuation on test-mass mirrors. These
  included two periodic signals mimicking a spinning neutron star in a
  binary system, similar to the low-mass X-ray binary
  \ac{ScoX1}. \ac{HW} injections serve as an important validation of a
  search method, as they occur before data cleaning and
  preparation. We searched for these injections using the model-based
  cross-correlation pipeline, which was used to search for \acp{CW} from \ac{ScoX1}
  in the first three LVK observing runs.
  To simulate a realistic search, we used a search
  region of orbital parameter space of similar size to the plausible
  parameter ranges for \ac{ScoX1}, but containing the parameters of
  each \ac{HW} injection. Our self-blinded analysis confidently
  detected one of the injections. The other, which was generated with a
  lower amplitude and in a more expensive region of frequency space,
  was not detected. We verified the imprint of this second signal by
  performing a deeper targeted search at its true parameters.
  However, such a deep search across the full parameter range would
  have been computationally infeasible.
\end{abstract}

\section{Introduction}

Low-mass X-ray binaries consist of a compact object (\ac{NS} or \ac{BH})
and a companion star. For a \ac{NS}, accreted
companion material can produce a mound on the surface, causing axial
asymmetry and generating \ac{GW} emission called {\acfp{CW}}.
The \ac{NS} in \ac{ScoX1}\cite{Giacconi1962_ScoX1, Bradshaw1999_ScoX1,
  Steeghs2002_ScoX1} is a \ac{CW} candidate of great priority, as its
high X-ray flux is indicative of a high accretion rate.
If the spinup and spindown torques are in equilibrium, the \ac{NS}
spin frequency, and therefore the \ac{GW} frequency $f_0$, can be
assumed to be approximately constant, but searches must consider the
residual uncertainty in orbital period ($\Porb$) and phase (written as
a reference time $\TascThen$), as well as the orbital speed $K_1$ or
equivalently the projected semimajor axis $a\sin i$ of the \ac{NS}
orbit.  Covering this parameter space presents a computational
challenge.
The model-based cross-correlation analysis
\cite{Dhurandhar2007_CrossCorr,Whelan2015_ScoX1CrossCorr} has
been used to conduct the most sensitive searches to date for \ac{ScoX1} in
simulated \cite{Messenger2015_MDC1} and actual
\cite{LVC2017_O1ScoX1CrossCorr, Zhang2021_O2ScoX1CrossCorr,
  LVK2022_O3ScoX1CrossCorr, Whelan2023_O3ScoX1NewEphem} \ac{GW}
detector data. In this work, we verify the effectiveness of the
pipeline by using it to search for simulated \ac{CW} signals called
\acf{HW} injections with parameters akin to \ac{ScoX1}.

\section{Hardware Injection Search}

\Ac{HW} injections are {simulated \ac{GW} signals} that are added to
LIGO data by perturbing the end mirrors \cite{Biwer2017_HWInj}. There were 18 \Ac{HW} injections in O3 (injections 0-17); two of them with parameters akin to \ac{ScoX1} (injections 16 and 17) \cite{LVK2023_O3OpenData}. Using parameter space coordinates defined in \cite{Wagner2022_Lattice}, we searched the O3 data for both injections 16 and 17 as a way to verify the effectiveness of the
cross-correlation pipeline in searching for real \ac{CW} signals.

\begin{table}[t!]
	\footnotesize
	\centering
		\begin{tabular}{cccc|cc}
			\hline
			 &  \multicolumn{2}{c}{\textbf{Search Region}} & & \multicolumn{2}{c}{\textbf{Injection 16 Results}} \\
			Parameter & \ac{ScoX1} & Injection 16 & Injection 17 & Recovered & Injected \\
			\hline
                  RA ($\alpha$) & $16^{\mathrm{h}}19^{\mathrm{m}}55.0850^{\mathrm{s}}$ & $1^{\mathrm{h}}19^{\mathrm{m}}55.0850^{\mathrm{s}}$ & $7^{\mathrm{h}}19^{\mathrm{m}}55.0850^{\mathrm{s}}$ & N/A & $1^{\mathrm{h}}19^{\mathrm{m}}55^{\mathrm{s}}$ \\
			Dec ($\delta$) & $15^\circ 38' 24.9''$ & $15^\circ 38' 24.9''$ & $15^\circ 38' 24.9''$ & N/A & $15^\circ 38' 24.9''$ \\
			Frequency (Hz) & unknown & $230-235$ & $890-895$ & $234.5670$ & $234.5670$ \\
			$a_p = a\sin i$ (lt-s) & $1.44-3.25$ & $1.43-3.21$ & $1.43-3.21$ & $2.3498$ & $2.3500$ \\
			$P_\text{orb} \pm 3.3\sigma$ (s) & $68023.92 \pm 0.056$ & $83941.19 \pm 0.056$ & $83941.19 \pm 0.056$ & $83941.2007$ & $83941.2000$ \\
			$\TascThen \pm 3\sigma$ (GPS) & $1078153676 \pm 99$ & $1036751085 \pm 99$ & $1036782289 \pm 99$ & N/A & N/A \\
$\TascNow \pm 3\sigma$ (GPS) & $1255015866 \pm 165$ &  $1254998190 \pm 165$ & $1255029394 \pm 165$ & $1254998237.1541$ & $1254998237.5020$ \\
			$K_1=\frac{2\pi a_p}{P_\text{orb}}$ (km/s) & $40-90$ & $32-72$ & $32-72$ & $52.7297$ & $52.7342$ \\
$h_0^\text{eff}$ & unknown & unknown & unknown & $1.09\times10^{-25}$ & $1.18\times 10^{-25}$ \\
			\hline
		\end{tabular}
                \caption{Search regions for \ac{HW} injections 16 and 17, compared with those used for \ac{ScoX1} in \cite{Whelan2023_O3ScoX1NewEphem}, along with the actual injected and recovered values for injection 16.  The sky position was assumed to be known; in fact the injected and search values of the right ascension were slightly different, since the former was rounded to the nearest second.  The difference is larger than the actual uncertainty in \ac{ScoX1}'s sky position, but smaller than the resolution of this search.  The ranges of orbital search parameters were chosen to be similar to those in \cite{Whelan2023_O3ScoX1NewEphem}.  In particular, the assumed uncertainties in orbital period $\Porb$ and time of ascension $\TascThen$ were propagated to the epoch of O3 to give the truncated elliptical search region in $\Porb$ and $\TascNow$ shown in Figure~\ref{followupratio}, using the methods of \cite{Wagner2022_Lattice}. $h_0^\text{eff}$, as defined in \cite{Messenger2015_MDC1} is the root mean squared of the plus and cross signal amplitudes.}
	\label{Parameters}
\end{table}

Instead of searching for the injections using their true parameters,
we used a parameter space region akin to that of
\cite{Whelan2023_O3ScoX1NewEphem}, using the PEGS~IV ephemeris from
\cite{Killestein2023_PEGS4} (see Table~\ref{Parameters}). We then
performed a hierarchical followup method, where we split the search
into four ``levels'', outlined in \cite{LVC2017_O1ScoX1CrossCorr}, to
filter out noise from true signal. {Level 0} is the initial search
with coherence time ${\Tmax}$ at threshold SNR ${\rho^\text{th}}$, and
we only keep candidates that have SNR $\rho > {\rho^\text{th}}$. The
threshold SNR is chosen to keep the number of false alarms manageable,
as detailed in \cite{Whelan2023_O3ScoX1NewEphem}.  {Level 1} is the
refinement of the search grid, increasing resolution. At {level 2}, we
perform the search again using this refined grid with $4\times$ the
original ${\Tmax}$. Lastly, at {level 3}, we perform the final search
with $16\times$ the original ${\Tmax}$. We invoke the condition to
veto any candidate whose SNR drops between consecutive levels; the SNR
for a true signal will increase with ${\Tmax}$, since searches with a
longer coherence time are more sensitive.

\section{Search Results and Analysis}

\subsection{Results}\label{Results}

Here we present the results of the searches for hardware injections 16 and 17, which were run on the Open Science Grid (OSG) \cite{osg06, osg07, osg09, osg15}.

\begin{figure}
	\centering
	\includegraphics[width=0.4\linewidth]{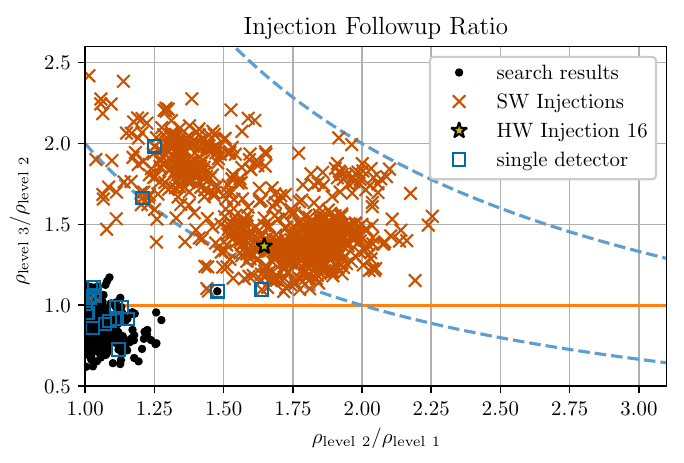}
	\includegraphics[width=0.42\linewidth]{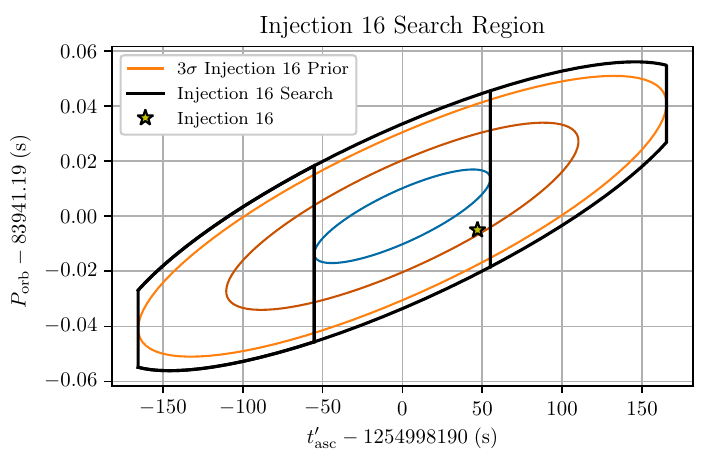}
	\vspace{-5mm}
	\caption{\textit{Left:} Followup ratio plot illustrating the SNR increase of digital \ac{SW} injections ($\times$) from \cite{LVK2022_O3ScoX1CrossCorr}, candidates from \cite{Whelan2023_O3ScoX1NewEphem} ($\bullet$), and injection 16 ($\star$). Injection 16 lies above $(1,1)$ in the coordinate plane (an expected signature of a true signal) and within the regime of digital software (SW) injections. Thus, injection 16 complements them. \textit{Right:} Search region for injection 16, including the $1\sigma$, $2\sigma$, and $3\sigma$ error ellipses. The ellipses, which contain injection 16, illustrate the assumed joint prior on the time of ascension and orbital period. This region is analogous to that shown in Figure~1 of \cite{Whelan2023_O3ScoX1NewEphem}.}
	\label{followupratio}
\end{figure}

\subsubsection{Injection 16}

Our initial search for injection 16 used the same parameters as in
\cite{Whelan2023_O3ScoX1NewEphem}.  The threshold \ac{SNR} was
${\rho^\text{th}}=6.1$, and coherence times of $2400$, $3000$, $4200$,
and $4800\un{s}$ were used in various regions of orbital parameter
space, as described in \cite{LVC2017_O1ScoX1CrossCorr,
  LVK2022_O3ScoX1CrossCorr}.  The signal was located in a region with
${\Tmax}=4200\un{s}$. At {level 0}, two candidates had
$\rho > {\rho^\text{th}}$.  Following the successive levels 1 and 2,
we found that $\rho$ decreased from {level 1} to {level 2} for one of
the candidates, meaning that it was only noise and was vetoed. The SNR
of the other candidate increased with ${\Tmax}$ for all levels, which
would mark it as a potential detection in a real search.  As shown in
Figure~\ref{followupratio}, the \ac{SNR} increase is consistent with
the \ac{SW} injections from \cite{LVK2022_O3ScoX1CrossCorr}, and
considerably larger than for any of the candidates from
\cite{Whelan2023_O3ScoX1NewEphem}.  Based on this, we conclude that
the search pipeline would have been able to detect this signal.  In
Table~\ref{Parameters} we list the parameter values inferred from the
last stage of followup, and compare them to the values with which
injection 16 was actually simulated.

\subsubsection{Injection 17}\label{Pulsar17}

The search for injection 17 used a threshold \ac{SNR}
${\rho^\text{th}}=5.6$ and coherence time ${\Tmax}=300\un{s}$.  (This
$\Tmax$ was used in
\cite{LVC2017_O1ScoX1CrossCorr,LVK2022_O3ScoX1CrossCorr,Whelan2023_O3ScoX1NewEphem}
because the search at $890-895\un{Hz}$ is inherently more expensive,
and the lower $\Tmax$ meant fewer templates and fewer loud outliers.)
Unlike injection 16, at
{level 0}, there were no candidates with $\rho >
{\rho^\text{th}}$. Therefore, the signal was too quiet to be detected with
${\Tmax}=300\un{s}$.
For reference, the injected parameters are tabulated in the Gravitational Wave Open Science Center (\href{https://gwosc.org}{gwosc.org})\footnote{Complete URL for tabulated parameters of hardware injections $0–17$: \url{https://gwosc.org/O3/O3April1_injection_parameters/}} \cite{LVK2023_O3OpenData}, and we compute $h_0^{\text{eff}}=6.21 \times 10^{-26}$. To determine the
necessary coherence time for a likely detection, we performed targeted
searches at the true parameters of injection 17 to obtain the SNR values
at ${\Tmax}=300, 1200$, and $4800\un{s}$. We compared these SNR values
to calculated expected SNR values and found
that injection 17 is, in principle, detectable using ${\Tmax}=4800\un{s}$. However, a full self-blind search
at this sensitivity in the $890-895\un{Hz}$ frequency band is
computationally infeasible, as we discuss in Section
\ref{Pulsar17CostAnalysis}.

\subsection{Cost Estimate for Injection 17 Detection}\label{Pulsar17CostAnalysis}
Since we were unable to detect injection 17 with $\Tmax=300\un{s}$, we
present a cost estimate for a search done at $\Tmax=4800\un{s}$. As
stated in Section \ref{Pulsar17}, a full self-blind search with this
coherence time would be expected to yield the detection of the
signal. We show here, however, that this is computationally
infeasible for a single $5\un{Hz}$ frequency band.

The cost of a search scales as the the product of the number of SFT
pairs and templates (points in parameter space):
$N_\text{Pairs}N_\text{Templates}$. This product roughly scales as
$N_\text{Pairs}f_0^2\Tmax^3$, so we use the injection 17 targeted
searches to estimate $N_\text{Pairs}$ and $N_\text{Templates}$ for a
blind search, as spacing values for the parameter space coordinates
are calculated as metadata for a specified mismatch. We estimate that
the cost of a blind search for $\Tmax = 4800\un{s}$ (in the
$890-895\un{Hz}$ band) is {$350553.0\un{CPU}\un{day}$}; over
$3.5 \times 10^4$ times more expensive than the search done at
$\Tmax=300\un{s}$.

\section{Conclusions}
Using the model-based cross-correlation pipeline, we searched for two continuous wave hardware injections (injections 16 and 17) that had parameters similar to that of Scorpius X-1. With search regions analogous to that in \cite{Whelan2023_O3ScoX1NewEphem}, we detected injection 16 (which had a frequency of $235.567\un{Hz}$), but were unable to detect injection 17 due to its lower amplitude and placement in a more expensive region of parameter space ($890-895\un{Hz}$ band). In principle, a search with a coherence time of $4800\un{s}$ (16 times longer than that of the original search) could recover the signal. However, we estimate that such a search has a cost of $~350553.0\un{CPU}\un{Day}$, which is computationally infeasible a single $5\un{Hz}$ frequency band.

\section*{Acknowledgments}

This work was supported by NSF grants PHY-2110460 and PHY-2409745 and has made use of data or software obtained from the Gravitational Wave Open Science Center (\href{https://gwosc.org}{gwosc.org}), a service of the LIGO Scientific Collaboration, the Virgo Collaboration, and KAGRA. This material is based upon work supported by NSF's LIGO Laboratory which is a major facility fully funded by the National Science Foundation, as well as the Science and Technology Facilities Council (STFC) of the United Kingdom, the Max-Planck-Society (MPS), and the State of Niedersachsen/Germany for support of the construction of Advanced LIGO and construction and operation of the GEO600 detector. Additional support for Advanced LIGO was provided by the Australian Research Council. Virgo is funded, through the European Gravitational Observatory (EGO), by the French Centre National de Recherche Scientifique (CNRS), the Italian Istituto Nazionale di Fisica Nucleare (INFN) and the Dutch Nikhef, with contributions by institutions from Belgium, Germany, Greece, Hungary, Ireland, Japan, Monaco, Poland, Portugal, Spain. KAGRA is supported by Ministry of Education, Culture, Sports, Science and Technology (MEXT), Japan Society for the Promotion of Science (JSPS) in Japan; National Research Foundation (NRF) and Ministry of Science and ICT (MSIT) in Korea; Academia Sinica (AS) and National Science and Technology Council (NSTC) in Taiwan. The authors additionally thank the OSG Consortium \cite{osg06, osg07, osg09, osg15}, which is supported by
the National Science Foundation awards \#2030508 and \#1836650, and by
the Digital Research Alliance of Canada (\url{https://alliancecan.ca}).
This paper has been assigned LIGO Document Number \dcc.

\providecommand{\newblock}{}

\end{document}